# Role of disorder and correlations in metal-insulator transition in ultrathin SrVO$_3$ films

Gaomin Wang[1], Zhen Wang[1,2], Meng Meng[1], Mohammad Saghayezhian[1], Lina Chen[1], Chen Chen[1], Hangwen Guo[1], Yimei Zhu[2], Ward Plummer[1], and Jiandi Zhang[1*]

[1]Department of Physics and Astronomy, Louisiana State University, Baton Rouge, Louisiana 70803, USA

[2]Brookhaven National Laboratory, Upton, New York 11973, USA

## Abstract

Metallic oxide SrVO$_3$ represents a prototype system for the study of the mechanism behind thickness-induced metal-to-insulator transition (MIT) or crossover in thin films due to its simple cubic symmetry with one electron in the 3$d$ state in the bulk. Here we report a deviation of chemical composition and distortion of lattice structure existing in the initial 3 unit cells of SrVO$_3$ films grown on SrTiO$_3$ (001) from its bulk form, which shows a direct correlation to the thickness-dependent MIT. *In-situ* photoemission and scanning tunneling spectroscopy indicate a MIT at the critical thickness of ~3 unit cell (u.c.), which coincides with the formation of a ($\sqrt{2}\times\sqrt{2}$)R45° surface reconstruction. However, atomically resolved scanning transmission electron microscopy and electron energy loss spectroscopy show depletion of Sr, change of V valence, thus implying the existence of a significant amount of oxygen vacancies in the 3 u.c. of SrVO$_3$ near the interface. Transport and magneto-transport measurements further reveal that disorder, rather than electron correlations, is likely to be the main cause for the MIT in the SrVO$_3$ ultrathin films.

*email: jiandiz@lsu.edu

**Introduction**

Transition metal oxides have been one of the most-studied systems in condensed matter research for many decades. The strong coupling between charge, spin, and lattice in these strongly correlated systems gives rise to a variety of fantastic phenomena such as high temperature superconductivity [1], colossal magnetoresistance [2], and multiferroicity [3]. Among the family of transition metal oxides, $SrVO_3$ (SVO) is a promising material that has recently attracted much attention for proposed applications as conducting electrodes [4] and transparent conductors [5]. Being a highly correlated paramagnetic metallic oxide, SVO is known to undergo metal-insulator transition (MIT) upon doping or reduced dimensionality in thin film form [6, 7]. The simple cubic structure and absence of magnetic ordering in SVO makes it a prototype for the study of the mechanism behind MIT, like structure distortion, electron correlations and disorder-induced localization.

Angle resolved photoemission spectroscopy (ARPES) study reported that MIT occurs at a critical thickness of 2-3 ML for SVO thin films grown on $SrTiO_3$ (001) (STO) substrate [6]. Based on Mott-Hubbard theory, MIT can be controlled by tuning the magnitudes of the on-site Coulomb repulsion *U* and bandwidth W [8, 9]. Going from bulk to thin films, the decrease in film thickness results in a reduction in the effective coordination number of constituent ions at the interface and surface, and therefore reducing the effective bandwidth W which may drive the system into MIT [6]. However, the change of lattice structure and/or chemical composition near the interface/surface may drastically alter the film properties. Recent studies with scanning tunneling microscopy (STM) and Spectroscopy (STS) [10,11] reported that the surface of SVO has two distinct surface termination associated with two surface reconstructions, a $(\sqrt{2}\times\sqrt{2})$-R45° and $(\sqrt{2}\times\sqrt{2})$-R26.5°, respectively. These reconstructions may be important factor to influence film metallicity in ultrathin limit, hence the observed MIT. However, the effect of film structure and chemical composition on MIT is neither studied nor understood. Furthermore, the disorder caused by defects such as oxygen vacancies and interface intermixing is almost unavoidable near the interface. STEM study [12] on SVO/STO demonstrated V-Ti inter-diffusion at interface and that interfacial defects induce a significant change in electronic properties by showing an electronic transformation from the insulating state to metallic state at SVO/STO heterointerfaces due to the hybridization of interfacial Ti 3d, O 2p and V 3d orbitals. It is proposed by the scaling theory of localization that in a 2D system, the existence of disorder alone will drive the system towards insulating behavior [13, 14], therefore disorder can play a crucial role in MIT. In the case of oxide thin films, disorder mostly exists in the form of oxygen vacancies, which can be easily introduced into the system during film growth at a low oxygen partial pressure or in vacuum. In addition, presence of epitaxial strain can promote the presence of the oxygen vacancies [15]. Therefore, in order to understand the origin of MIT, the role of disorder (cation or anion) must be considered on the same grounds as electron-electron correlations. In this study, we seek to understand the role played by disorder behind the MIT in SVO system, by investigating the structure and electronic properties of the SVO thin films with various tools including ultraviolet photoemission spectroscopy (UPS), scanning tunneling spectroscopy (STS), scanning transmission electron microscopy (STEM) and electron energy loss spectroscopy (EELS). Our study confirms that near the SVO/STO interface where the film is under maximum strain, there indeed exist a significant amount of cation and anion defects. Moreover, by introducing more oxygen vacancies into the original metallic SVO thin films, one can eventually drive the films to insulating. We therefore conclude that disorder plays a significant role in the thickness-induced MIT.

**Experimental Details**

The substrates used in this study were 0.1% Nb: $SrTiO_3$ (001) samples supplied by Crystec GmbH. Before transferred to the vacuum chamber, every substrate was cleaned with acetone, alcohol and deionized water, etched with Buffered HF for 30 seconds and subsequently annealed at 950 °C for 1h in oxygen to prepare an atomically flat $TiO_2$-terminated surface for an optimized film growth. The SVO films were grown at substrate temperature of 600 °C under vacuum level of $1\times10^{-8}$ Torr with laser energy of ~2.5 $J/cm^2$ and monitored by reflection high-energy electron diffraction (RHEED) to guarantee a layer-by-layer 2D growth. A stoichiometric SVO target was used for the deposition. After the growth, the films were transferred to a characterization chamber through vacuum to perform *in-situ* analysis. Scanning tunneling microscopy/spectroscopy (STM/STS) was taken with a variable temperature STM from Omicron. UPS was measured with a SPECS PHOIBOS-150 analyzer calibrated with gold single crystal, with a monochromated Scienta VUV 5000 Helium photon source. LEED from the surface of SVO films were also performed *in-situ*. Cross-section TEM sample were prepared by Focused Ion Beam (FIB) with $Ga^+$ ions following by the $Ar^+$ ions Nano-milling. The microscopy work was performed *ex-situ* on a

JEOL ARM200 microscope at Brookhaven National Laboratory (BNL) equipped with two aberration correctors. High-angle annular dark-field (HAADF) STEM images were collected with a HAADF detector with an inner angle of 67 mrad and outer angle of 275 mrad. All electron energy loss spectroscopy (EELS) were acquired at a semi-convergence angle of 20 mrad and a semi-collection angle of 88 mrad. An energy dispersion of 0.25eV/channel with energy resolution of ~0.8eV was used for fine structure study of EELS spectra. The EELS spectra of V-$L_{2,3}$, Ti-$L_{2,3}$, O-K and Sr-$L_{2,3}$ edge was obtained simultaneously at an energy dispersion of 1eV/channel for composition analysis. Dual EELS mode was used in order to calibrate the energy shift in the spectra collection process. The resistivity and magnetoresistance were measured by a physical properties measurement system (PPMS) from Quantum Design Inc. The samples for all *ex-situ* measurements were capped with amorphous STO for the protection of SVO films.

**Results and Discussions**

A ball model of the SVO/STO structure is displayed in Fig. 1(a). The bulk of both SVO and STO have cubic structure, with lattice constant of 3.842 Å and 3.905 Å, respectively, giving the film a tensile strain with a lattice mismatch of approximately 1.6%. Fig. 1(b) shows the RHEED patterns and the corresponding oscillations for a 20 u.c. SVO film, which indicates layer-by-layer 2D growth. STM morphology images and height profiles of two SVO films of thickness 3 u.c. and 50 u.c. are shown in Figs. 1(c-d). For the 3 u.c. film, a good film growth following the terraces of the STO substrate is clearly indicated; as the film grows thicker, the terrace edges are no longer distinguishable. However, the surface roughness of the 50 u.c. film is still maintained at the height of one single unit cell, as in the 3 u.c. film. The absence of small clusters on the film surface from the STM image confirms that our SVO films are fabricated in the 2D layer-by-layer fashion with atomically flat surfaces.

To confirm the thickness-dependent MIT, we have grown SVO films with different thicknesses and performed the measurements of the density of states near Fermi energy with UPS and STS, which are shown in Fig. 2. A drop of the intensity at the Fermi edge is clearly noticeable in UPS curves of the SVO films with decreasing thickness, which eventually evolves into a gap with the valence band edge at ~ 0.4 eV below $E_F$ at the film thickness of 1 u.c., as shown in Fig. 2(a). The detailed intensity v.s. film thickness plot at the Fermi edge is shown in Fig. 2(b). Given the error bar of the measurements, we can see that below 3 u.c. thickness, there is no intensity at the Fermi edge. Starting from 3 u.c., the intensity at the Fermi edge starts to increase with increasing film thickness. The intensity at the Fermi edge for 4 u.c. film almost reach to the same level for 25 u.c. one, indicating that the film becomes fully metallic, which is consistent with the results previously reported elsewhere [6].

The MIT at a critical thickness of 2~3 u.c. is also confirmed by the STS measurements. Although STS allows to access local density states, by taking measurements on a reasonable number of points randomly sampled in different locations on the same sample, one can still compare and average the results to obtain a more general look of the film's electronic property. In our measurements of tunneling current as a function of electric bias voltage (*I-V*) curves, ten or more points are sampled for each film, with about 20 curves taken for each point. Fig. 2(c) presents the averaged *I-V* curves of SVO films of different thicknesses compared with the one from STO substrate, while $dI/dV$ as a function of electric bias voltage curves obtained by lock-in measurements of the *I-V* curves are displayed in Fig 2(d). With the directly proportional relationship between $dI/dV$ and local density of states, based on STS theory [16], $dI/dV$ curves provide us another indication of the metallicity of the sample. From the $dI/dV$ results, obvious gaps can be observed in the both STO substrate and 1 u.c SVO film; in the 2 u.c. film the gap is clearly closing and yields zero density of states only at zero bias; as the film thickness further increases, the density of states near zero bias increases, completing the MIT.

To understand the underlying mechanism of this thickness-dependent MIT, we have performed the structural characterization of SVO films with both LEED and STEM. As shown in Fig. 3(a), LEED image appears $p(1\times1)$ pattern for the surface of a simple cubic-like lattice. However, fractional spots appear for the films with 3 u.c. and above as shown in Fig. 3(a) and the line profile of LEED intensity for different film thickness shown in Fig. 3(b). The LEED pattern with fractional spots indicates a $(\sqrt{2}\times\sqrt{2})R45°$ surface reconstruction. Such a surface reconstruction has been observed previously by LEED and STM image from thick SVO films [10, 11]. STM image seems to indicate that the superlattice surface structure is formed by a $VO_2$-terminated layer with apical oxygen adsorption. A 50% coverage of the apical oxygen sites (forming a $SrVO_{3.5}$ surface structure) has been suggested [10, 11], which leads to the formation of the $(\sqrt{2}\times\sqrt{2})R45°$ LEED pattern. Interestingly, the formation of such ordered structure coincides with the metallization of films.

We performed STEM/EELS measurements to study atomic structure and chemical composition across the SVO/STO interface. We have chosen an SVO film with a thickness of 50 u.c. in this study to ensure the interface region is protected from degradation in the atmosphere. Figure 4(a) shows the low magnification cross-section Z-contrast HAADF-STEM image across the SVO/STO interface taken along [100] direction. Interestingly, there is a dark-line appearing along the interface. Figure 4(b) presents an atomically-resolved image near the interface. As marked by the red dotted line, the STO substrate terminates with $TiO_2$ layer. The sharp interface indicates coherent epitaxial growth of the SVO film on the STO substrate. The column intensity profile curve as shown in Fig. 4(b) shows that there is a clear intensity depletion, *especially the brightness of Sr sites*, in the first 3 u.c. SVO film from the interface (i.e., the area from the yellow-dashed line to the interface). For convenience, we will refer the first 3 u.c. of SVO film as the "*dark area* (DA)". This corresponds to the dark-line near the interface observed in Fig. 4(a). The column intensity in HAADF-STEM image is proportional with atomic number ($Z^{1.7}$) [17], and the column intensity decrease from Sr ($Z = 38$), V ($Z = 23$), to Ti ($Z = 22$). Therefore, the depletion of intensity at Sr-sites in the Z-contrast HAADF-STEM image show a deficiency of Sr in the first 3 u.c. of SVO film. On the other hand, the SVO film has a simple cubic-like structure, i.e. a tetragonal structure. This is also evident from the diffraction pattern shown in Fig. 4(c), taken along [100] direction from the SVO film and STO substrate. As displayed in Fig. 4(d), the annular bright-field (ABF)-STEM image (which is sensitive to oxygen sites) shows no obaservable tilt distortion of $VO_6$ octahedra in SVO film, consistent with cubic-like structure.

However, the first 3 u.c. of SVO (the DA region) shows different lattice constant in the *c*-axis and chemical composition from the rest of the film. With respect to each atomic layer in the HAADF-STEM image presented in Fig. 5(a), we plotted in Fig. 5(b) the out-of-plane lattice constant as a function of distance from the interface, averaged over 15 u.c. along the [010] direction. For SVO film beyond the DA, the measured out-of-plane lattice constant converges to the value of bulk SVO (3.84 Å, see the dotted line). Yet, the out-of-plane lattice constant expands to 3.97 ± 0.4 Å in DA, larger than bulk STO (3.905 Å) which is not expected in a film under tensile strain. The abnormal lattice expansion (so as the unit cell volume) in the DA is likely related to existence of considerable oxygen and strontium deficiency. There have already been reports of oxygen-deficiency driving the lattice expansion in complex oxides, both in bulk and in thin film [18-20].

Also, the fact that the DA is less contrasted in HAADF-image itself suggests possible existence of oxygen-deficiency [21].

We explored the film further by performing EELS elemental mapping to characterize the composition distribution and electronic structure across the interface. Figures 5(c-e) show atomic resolved elemental mappings of Sr-$L_{2,3}$, V-$L_{2,3}$ and Ti-$L_{2,3}$ edges, respectively. EELS composition profile can be obtained by averaging along the interface. The elemental concentration profiles are derived from the intensity profile, which is plotted on the corresponding atomic sites in the elemental maps. The STO substrate terminates with a $TiO_2$ layer as expected from the substrate preparation. Assuming the Ti and V level to be 100% in STO and SVO far away from the interface, respectively, the Ti intensity drops to ~ 80% in the $TiO_2$ termination layer of the substrate, ~ 30% in first $VO_2$ layer and ~ 10% in the second $VO_2$ layer. This suggests the existence of V-Ti intermixing at B-sites within the first 1-2 u.c. of the SVO film and the top layer of substrate. A ~ 10% depletion of Sr concentration is also observed within the first 3 u.c. SVO from the interface. The EELS fine structure is related to the details of the unoccupied states. The EELS spectra of V-$L_{2,3}$ and Ti-$L_{2,3}$ edge taken across the interface are displayed in Figs. 5(g-h). The extraction of O-K edge in SVO is complicated by the proximity of the V $L_{2,3}$-edge. For better comparison, the EELS spectra have been normalized to the integrated intensity under the V L-edge; the spectra of Ti L-edge are normalized to the continuum interval 25eV before the onset of the oxygen K-edge. A red shift of ~ 0.4 eV of the V $L_3$-peak is observed within the first 3 u.c. from the interface (i.e., the DA), indicating a reduced V valence state [12]. The two spectra of the $TiO_2$ termination layer and the doped $VO_2$ layer also shift towards lower energy, implying decreased valence state of Ti ions [22]. The details on the valence change of V and Ti are plotted in Fig. 5(f). A drop of V-valence to about ~ 3.6+ within the DA is evident. The oxidation state of Ti decrease to ~ 3.5+ in the DA. As a system which does not favor tilt and rotation and being non-polar along the [100] direction, no polar-discontinuity need to be compensated by structural/charge reconstruction at the interface when SVO is grown on STO. In order to maintain charge neutrality (in DA) near the interface, a significant amount of oxygen-deficiency is suggested; one part for compensating the slight-Sr deficiency, and other part for the reduced oxidation of Ti and V. Recalling that critical thickness of metal-insulator transition for SVO films is also 3 u.c., this leads to our hypothesis that the disorder due to oxygen as well as strontium deficiency may be the main driving force for the MIT.

As mentioned in the introduction, disorder alone, in the form of oxygen deficiency, can drive our system into insulating. If this is the case, then it is quite possible that we should be able to tune a metallic SVO film to insulating by introducing more oxygen deficiency into the film on purpose. We tested our hypothesis by post-annealing the SVO films in the same vacuum level for film growth but under higher temperature. *Ex-situ* measurements of resistance and magnetoresistance were performed on the post-annealed films along with a set of films of different thicknesses for comparison. Figure 6(a) displays the sheet resistance measurements of SVO films with thicknesses from 3 u.c. to 20 u.c.. The 3 u.c. film clearly shows insulating behavior, while the other films exhibit MIT or metal-insulator crossover. For 3 u.c. film, the *T*-dependence of the resistance is understood with the 2D variable-range hopping (VRH) model due to strong localization [23]. The 4-10 u.c. SVO films exhibit metallic behavior at high temperature, while an upturn of sheet resistance with decreasing temperature is observed in the low temperature region, indicating the existence of weak localization. The temperature for the resistance minimum increases with decreasing film thickness, reflecting the enhancement of disorder-induced localization strength with reducing thickness, as observed in several oxide films [24-26]. In Fig. 6(b), we converted the sheet resistance to conductance and plotted the curves with respect to the logarithm of temperature. In all four films with the 4-8 u.c. films, good linear fits of sheet conductance v.s. ln*T* were obtained in the low temperature region, which is consistent with the Anderson localization scenario though at the same time unable to rule out the contribution of electron-electron correlations. To further gain insight the disorder effect on transport due to the oxygen vacancies, we have carried out the transport measurements for a 10 u.c. film by different vacuum-post-annealing time which introduces different vacancy concentration and the results are shown in Fig. 6(c). The 10 u.c. film measured as grown shows metallic behavior. When post-annealed under 850 ˚C for 20 minutes, the film remained metallic; while as the annealing time increased to 40 minutes, the film began to act insulating. This shows that by introducing more oxygen deficiency, we are indeed able to drive the originally metallic SVO film to insulating.

To further understand the nature of localization-induced insulating behavior, we have also measured the magnetoresistance (MR) for a set of SVO films, including the 10 u.c. films with and without post-annealing. It has been predicted by theory that disorder-induced localization leads to negative MR [27-29] while electron-electron correlations lead to positive MR [30]. In Figs. 7(a-c), we presented the MR of 4, 6, and 10 u.c. SVO films under different temperature. At 5K, the

MR for the 4 u.c. film is negative, indicating the dominance of disorder-induced localization; while for the 10 u.c. film, this negative MR effect is no longer visible. Instead, positive MR emerges although the value of MR is almost 10 times smaller than that of 4 u.c. film. However, after the vacuum-post-annealing treatment that introduces more oxygen vacancies, the MR for the 10 u.c. film started to turn negative again, as shown in Fig. 7(d) and compared with Fig. 7(c). Combined with our transport results, we can conclude that disorder-induced localization is responsible for insulating behavior at low temperature for the ultrathin films. Such disorder effect is amplified with reduced film thickness or vacuum-post-annealing, based on the enhancement of negative magnetoresistance. With increasing temperature or film thickness (see at $T$ = 9 K and 13 K, for example), the magnitude of negative magnetoresistance drastically decreases while the weak positive magnetoresistance remains. The positive magnetoresistance, which is likely due to the contribution of electron correlations [30], does not strongly depend on the film thickness in such ultrathin film case. Therefore, disorder effect, which is enhanced with reducing film thickness, is the major driving force behind the MIT.

**Conclusion**

We have grown SVO films in a layer-by-layer fashion and show the thickness-dependent metal-insulator transition with a critical thickness of ~3 u.c. by UPS and STS. The structural, compositional, and oxidation state evolution near the SVO/STO interface were studied by atomic resolved HAADF-STEM and STEM-EELS. The reduction of V oxidation state within first 3 u.c. of SVO from the interface was observed. Combining it with Sr-depletion within the same region, we conclude that significant oxygen deficiency occurs within the first 3 u.c. of SVO. From the transport and MR measurements of the SVO films with and without vacuum-post-annealing, we have concluded that disorder-induced localization is the main driving force for the MIT in the SVO thin films, although electron correlations also exist in the system.

*Acknowledgments*: This work was primarily supported by U.S. DOE under Grant No. DOE DE-SC0002136. G.W. was supported by U.S. NSF under Grant No. DMR 1608865. The electronic microscopic work done at Brookhaven National Laboratory is sponsored by the US DOE Basic Energy Sciences, Materials Sciences and Engineering Division under Contract DE-AC02-98CH10886.

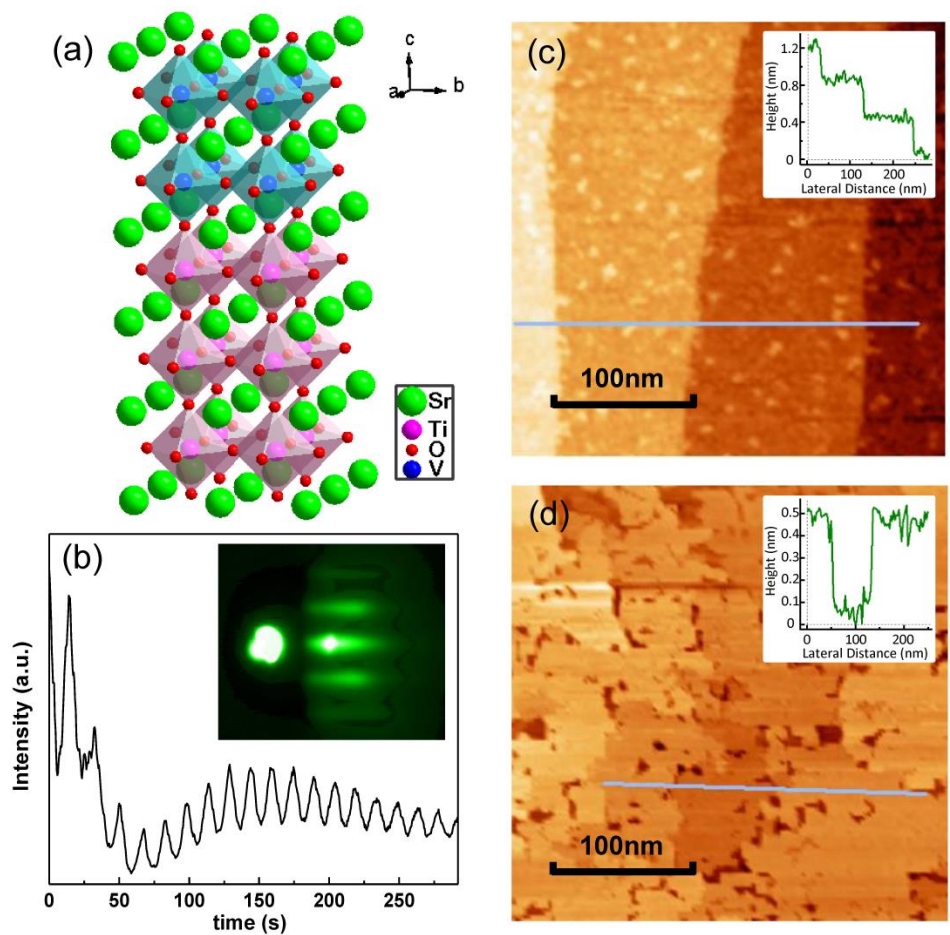

**Fig. 1** (a) A structural model of SrVO$_3$ film grown on SrTiO$_3$ (001). (b) The reflection high-energy electron diffraction (RHEED) intensity oscillation curve and pattern of a 20 u.c. SrVO$_3$ film. (c)-(d) The scanning tunneling microscopy (STM) images of a 3 u.c. and a 50 u.c. SrVO$_3$ film, respectively, with line profiles shown in the insets. The size of the images is 300 nm x 300 nm.

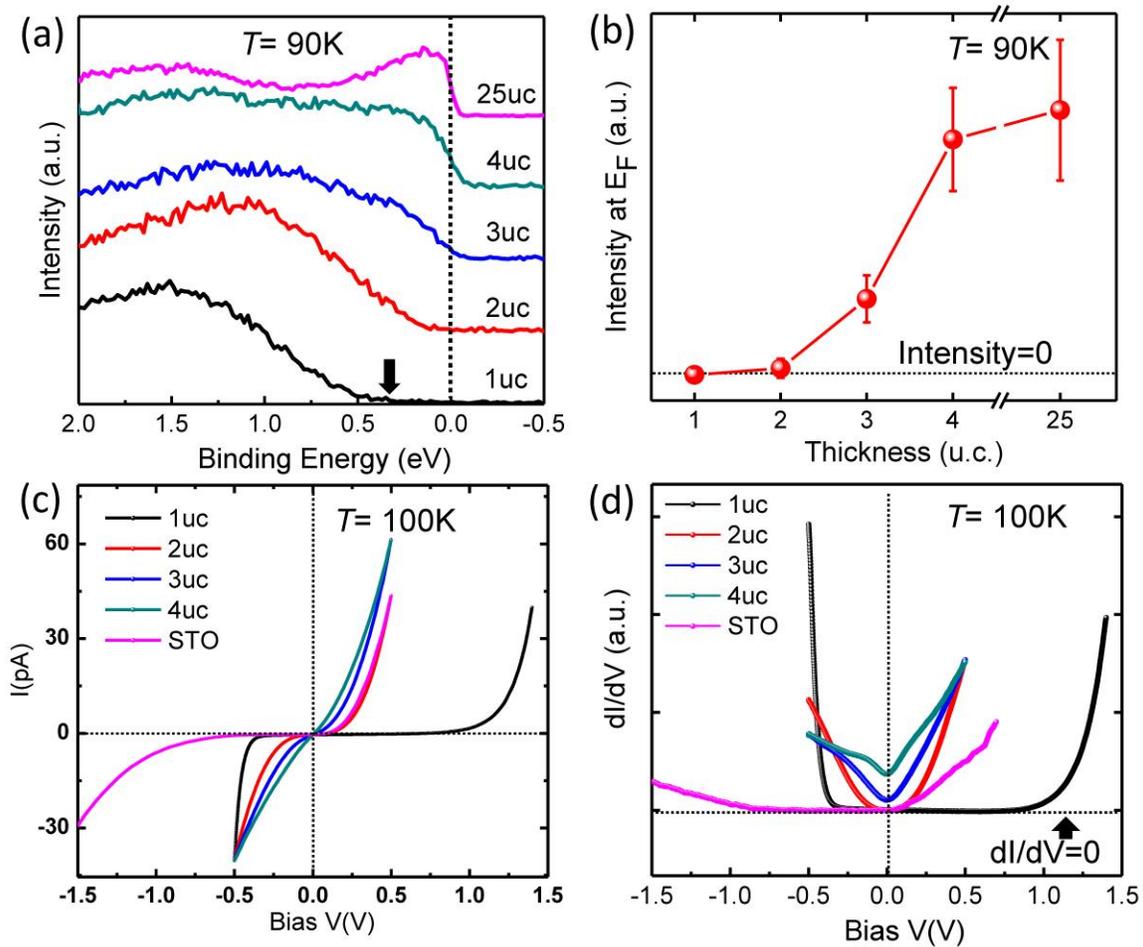

**Fig. 2** (a)-(b) The ultraviolet photoemission spectra (UPS) of SrVO$_3$ films with thicknesses of 1-4 u.c. and 25 u.c. near the Fermi edge ($E_F$), and the corresponding intensity change at the Fermi edge with increasing thickness. (c) Scanning tunneling spectroscopy *I-V* curves for 0.1% Nb-doped SrTiO$_3$ substrate and SrVO$_3$ films with different thicknesses. (d) (d*I*/d*V*)-*V* curves for the samples in (c).

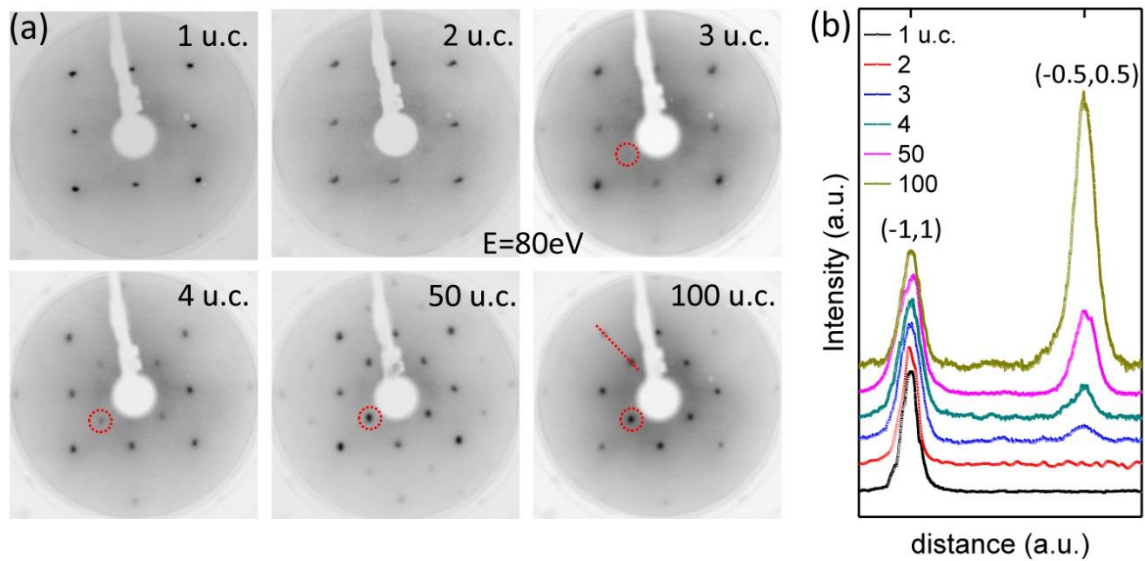

**Fig. 3** Low energy electron diffraction (LEED) patterns from the surface of SrVO$_3$ films with thicknesses of 1-4, 50 and 100 u.c. with electron beam energy E = 80 eV. (d) Intensity profiles along the cut line across the integer spot (-1,1) and the fractional spot (-0.5,0.5) from the patterns displayed in (c). The red dotted line is shown in the 100 u.c. pattern in (c) as the red dotted line. The red circles mark the fractional spots.

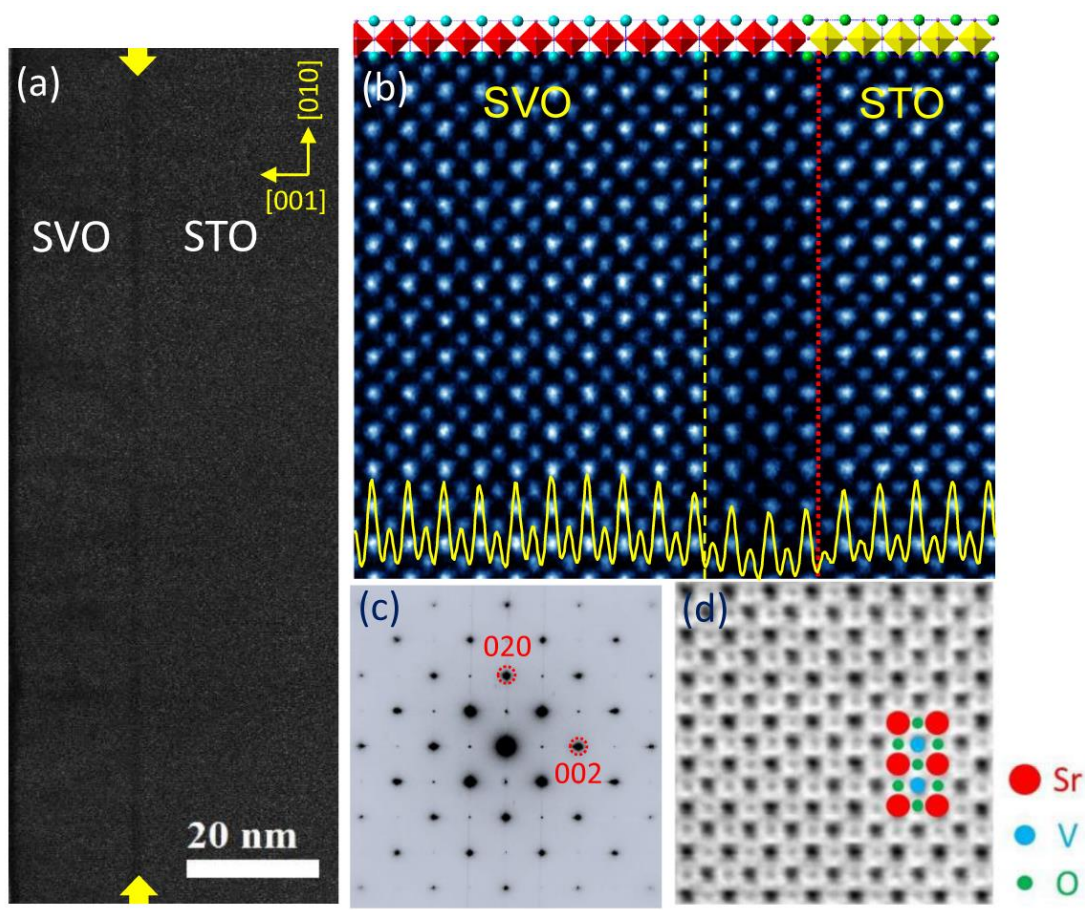

**Fig. 4** (a) Low magnification HAADF-STEM image taken along [100] direction across the interface of a 50 u.c. SVO film on STO(001). The yellow arrows indicate the darker region in the SVO near the interface. (b) Enlarged HAADF-STEM image with intensity profile superimposed and ball-model of the structure. The red dotted line marks the position of interface and yellow dotted line the SVO layers with lower intensity from the interface. (c) Selected area diffraction pattern taken along [100] direction from the SVO film and STO substrate. (d) ABF-STEM image of the SVO film with structure model superimposed.

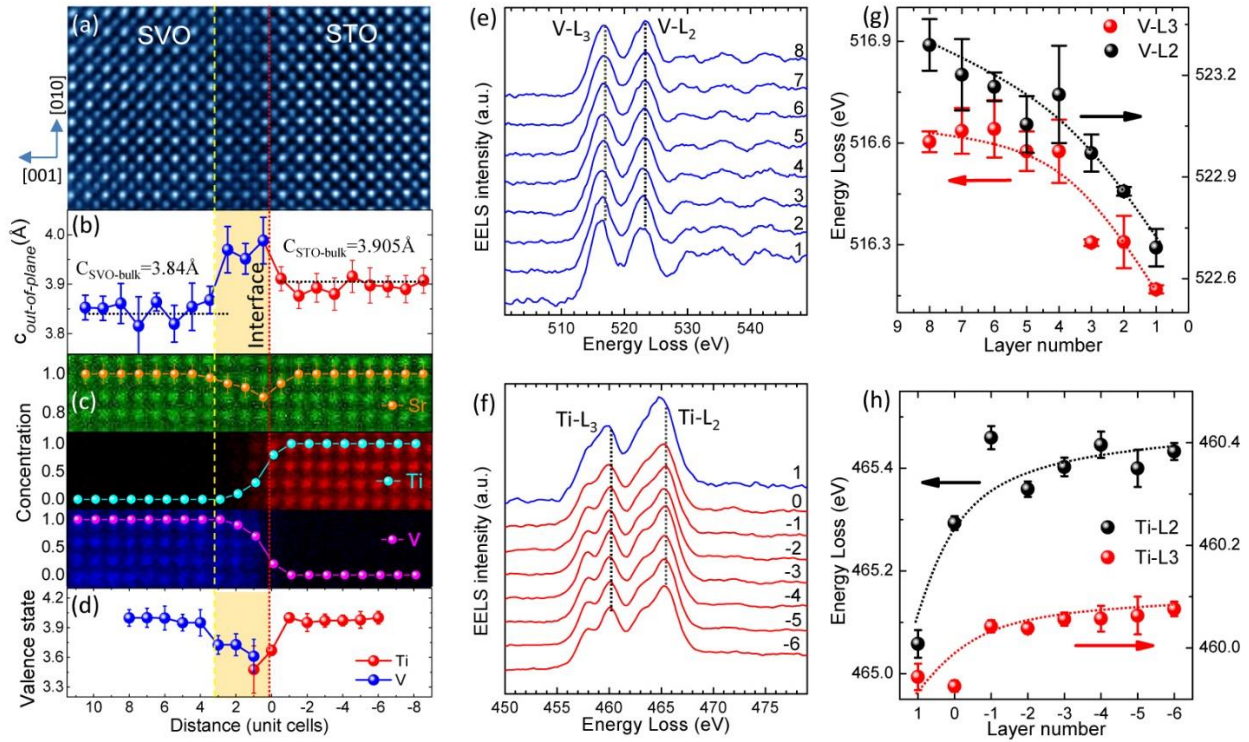

**Fig. 5** (a) HAADF-STEM image across the 50 u.c. SVO/STO film taken along [100] direction. (b) Out-of-plane lattice constant as a function of distance from the interface (x=0), measured from HAADF-STEM image by averaging 20 u.c. along the [010] direction. The lattice constant for bulk STO and SVO are indicated by dotted lines. (c) Falsed colored elemental maps for Sr (green), Ti (red) and V(blue), with lateral averaged profiles overlaid. The red dotted line marks the interface and the off-stoichiometry region of the film is highlighted. (d) Oxidation state of Ti and V ions across the interface. (e-f) Background subtrated V-$L_{2,3}$ and Ti-$L_{2,3}$ EELS spectra. (g-h) The energy of V-L and Ti-L edge peaks (The dashed lines are quide to eyes).

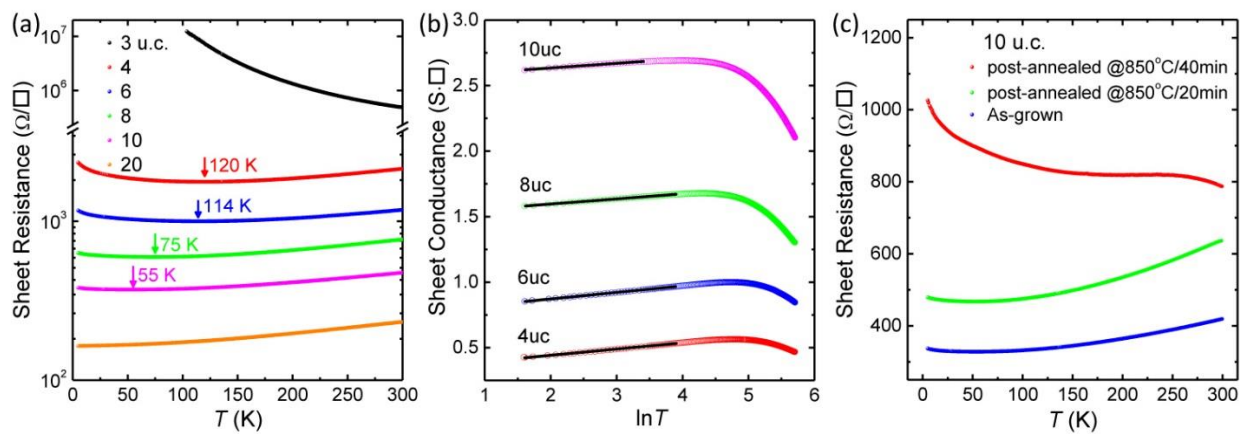

**Fig. 6** (a) $T$-dependence of sheet resistence of SrVO$_3$ films with different thicknesses. The arrows mark the resistance mimima. (b) The sheet conductance of SrVO$_3$ films with different thicknesses, fitted to the logarithm of the temperature in the low-$T$ region (5-30K for the 10 u.c. film, and 5-50 K for the 4-8 u.c. films). (c) $T$-dependence of sheet resistence of 10 u.c. SrVO$_3$ films as grown and post-annealed under 850 ˚C in high vacuum for different time.

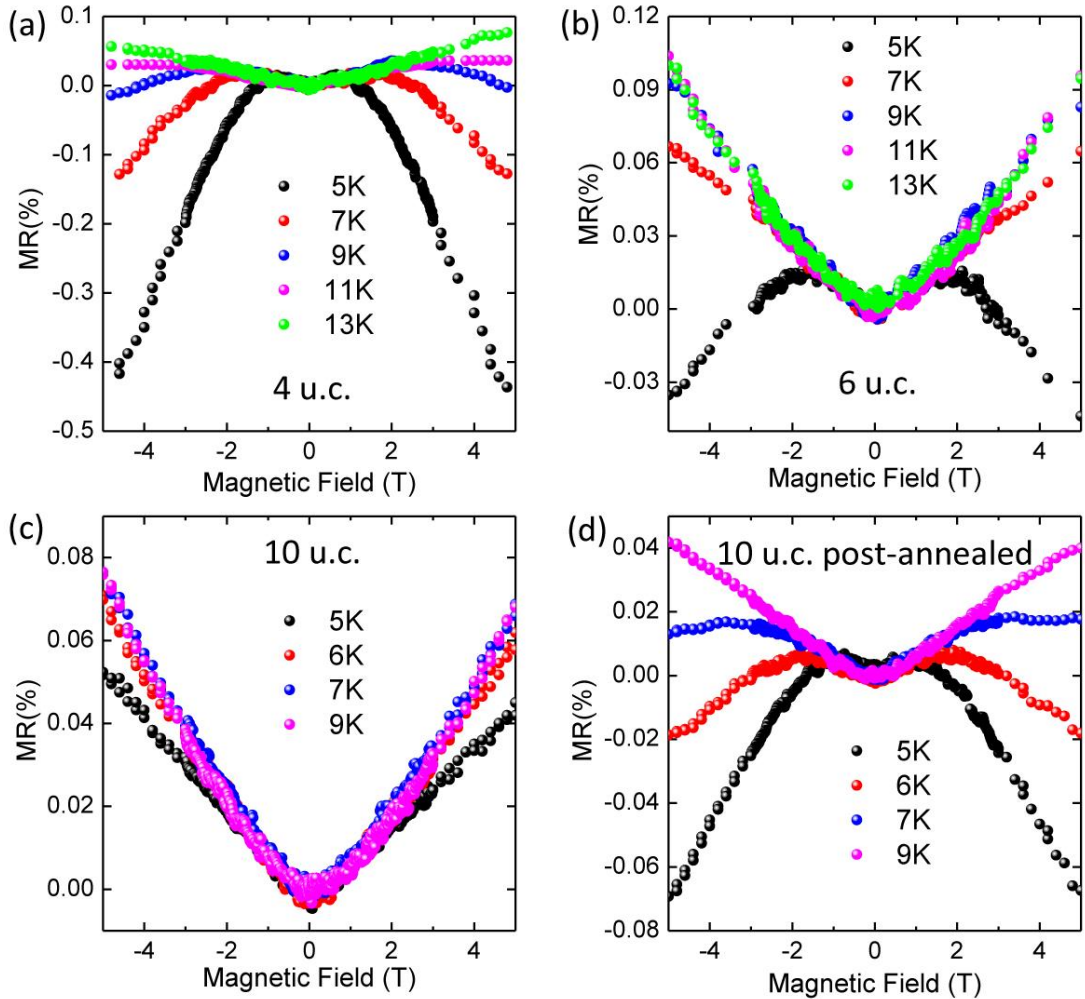

**Fig. 7** Magnetoresistance (MR) in magnetic field perpendicular to the SVO films at different temperature and with the thickness of (a) 4 , (b) 6, (c) 10, and (d) 10 u.c. post-annealed in high vacuum under 850 ˚C for 40 minutes.